\documentclass[aps,preprint,groupedaddress,a4paper]{revtex4-2}

\usepackage[T1]{fontenc}   
\usepackage{textcomp} 
\usepackage{graphicx}

\usepackage{mathtools}
\usepackage[colorlinks = true,linkcolor = blue,urlcolor  = blue,citecolor = blue,anchorcolor = blue]{hyperref}
\usepackage{xcolor, soul}
\usepackage{color}

\usepackage{makecell}

\DeclarePairedDelimiter\ket{\lvert}{\rangle}
\DeclarePairedDelimiter\bra{\langle}{\rvert}

\begin{document}


\title{Fault-Tolerant Encoding of Logical Qudits in Spin Systems}


\author{Sumin Lim}
\email[]{suminlim@kaist.ac.kr}
\affiliation{Graduate school of quantum science and technology, KAIST, Daejeon, 34141, Republic of Korea}
\affiliation{CAESR, Department of Physics, University of Oxford, The Clarendon Laboratory, Parks Road, Oxford OX1 3PU, UK}



\date{\today}

\begin{abstract}
Universal quantum computers require fault-tolerant logical qudits, as qudits naturally align with the simulation of multi-level physical systems. Here, we present a general framework and working examples for encoding fault-tolerant logical qudits in finite-dimensional spin systems. We construct distance-$3$, distance-$5$ codewords, and general $2t+1$-distance codes that can be implemented using a single physical qudit or a small number of coupled qudits for higher distances, while requiring a Hilbert space dimension significantly smaller than conventional constructions based on multiple logical qubits. Logical operations and error correction protocols can be implemented with polynomial scaling in the number of elementary operations. We further discuss schematic designs for physical implementation and required single-gate fidelities, which are compatible with current spin qudit platforms. This strategy provides a resource-efficient path toward realizing fault-tolerant logical qudits in finite multi-level physical systems. 

\end{abstract}


\maketitle

\section{Introduction}

Quantum technologies -- including quantum information processing\cite{shor1999polynomial, vandersypen2001experimental,grover1996fast, chuang1998experimental, cory1997ensemble}, quantum simulation\cite{Feynman1982Simulating,lloyd1996universal,aspuru2005simulated}, quantum sensing\cite{degen2017quantum, dolde2014nanoscale, clarke2008superconducting}, and quantum communication\cite{bennett2014quantum} -- have attracted significant interest across diverse research communities due to their potential to outperform classical approaches. Among these, quantum simulation provides a powerful framework for studying complex physical systems by directly emulating their intrinsic quantum properties, as originally proposed by Feynman\cite{Feynman1982Simulating}. While qubit-based architectures can represent multi-level quantum states, they typically requires additional physical qubits and increased circuit depth. A more natural and resource-efficient approach is to directly exploit qudits, i.e., quantum systems with intrinsic multi-level structure. Recent theoretical and experimental studies \cite{chizzini2024qudit, campbell2014enhanced, chizzini2022molecular} have explored various kinds of qudit-based quantum simulations, including applications to molecular dynamics\cite{macdonell2021analog}, neutrino oscillations \cite{turro2025qutrit}, and disorder-induced superfluidity\cite{ticea2025observation}, highlighting the potential of quantum multi-level simulators. Beyond quantum simulation, qudit-based quantum information processing \cite{lim2023fault,jankovic2024noisy,gao2023role,sawaya2020resource} and quantum communication \cite{bechmann2000quantum, cerf2002security, cozzolino2019high} inherently offer enhanced information capacity and improved computational or communication efficiency, thereby motivating the development of dedicated qudit-processing schemes. 

It is widely recognized that we are currently in the noisy intermediate-scale quantum (NISQ) era\cite{preskill2018quantum}. In this context, it is essential to develop strategies for encoding logical qudits using physical qudits\cite{uy2025qudit, brock2025quantum, michael2016new, ouyang2017permutation, mazurek2020quantum, schmidt2022quantum, dutta2025noise}, analogous to the encoding of logical qubits in physical qubits\cite{knill1998resilient, laflamme1996perfect, knill1997theory, google2023suppressing}. Although the Gottesman-Kitaev-Preskill (GKP) code \cite{gottesman2001encoding} intrinsically implies the encoding of logical qudits, it relies on bosonic systems\cite{hu2019quantum, brock2025quantum}, i.e., an infinite-dimensional Hilbert space, along with additional state-normalization procedures and corresponding experimental overhead. An alternative approach is to exploit intrinsically finite-dimensional multi-level systems, such as spin qudits. 

Spin systems, which were proposed and demonstrated as candidate embodiments in the early stages of the quantum computing era\cite{vandersypen2001experimental, cory1997ensemble, chuang1998experimental}, can also provide a robust multi-level platform. Electron spins offer fast gate operations and inter-qubit interactions across various platforms, while nuclear spins are known to be one of the most stable quantum systems, and are promising candidates for fault-tolerant quantum memory. In recent years, extensive investigations have been performed to exploit spin qudit systems as quantum memory, in both experimental\cite{lim2025demonstrating, yu2025schrodinger, yang2025minute} and theoretical\cite{gross2021designing, lim2023fault, omanakuttan2023multispin, omanakuttan2024fault, omanakuttan2023qudit, chiesa2020molecular, lockyer2021targeting, carretta2021perspective, lim2025designing} studies. This approach is not limited to spins in solid-state systems\cite{fernandez2024navigating, takahashi2009coherent, liu2018qubit}, but can also be extended to encodings in angular-momentum space in trapped and laser-cooled atoms or molecules\cite{albert2020robust, jain2024absorption}, which can have higher total angular momentum\cite{boguslawski2023raman, yu2022magneto} than typical nuclear spin systems, as well as to photonic platforms exploiting orbital angular momentum degrees of freedom\cite{yao2011orbital, kim2024qudit}. 

In this study, inspired by various qudit-based codes\cite{uy2025qudit, gross2021designing, lim2023fault, omanakuttan2023multispin} including \textit{cat} codes\cite{leghtas2013hardware, ofek2016extending} and \textit{binomial} codes\cite{ni2023beating, li2021phase, michael2016new}, we propose a method for encoding a logical qudit using either a single larger spin qudit or multiple spin qudits, with only polynomial scailing in Hilbert space dimension and gate complexity. We first present working examples of logical qutrit and ququart encodings against first-order Z errors (phase errors), and their generalization to arbitrary $d$-dimensional qudits. We next extend the codewords to enable correction of Pauli X, Y, and Z error correction, and describe how the design can be generalized to $d$-dimensional qudits. Additionally, this strategy can be expanded to second- and higher- order error corrections, and we provide brief steps for its extension. Depending on the system, multiple coupled spin qudits can be entangled to form a fault-tolerant logical qudit, with efficient encoding-decoding schemes, which we illustrate with working examples. Finally, we provide a discussion of schematic design for the physical implementation, fault-tolerant logical qudit operations, and required single gate fidelities. The results clearly demonstrate that direct encoding of logical qudit provides an advantage in quantum resource efficiency.

\section{$Z$-error correction code for qutrit, ququart, and generalization}
\label{Chapter_Zerror}

In this section, we outline a general formalism for constructing logical qudits with arbitrary dimension $d$ and codeword distance $2t+1$. The logical codewords are constructed as linear combinations of symmetric \textit{cat} -like superpositions within the $Z$ basis, with coefficients carefully chosen to satisfy the qudit Knill-Laflamme conditions up to order $2t$. For distance-3 $Z$ error correction, satisfying first-order KL constraint requires at least two independent \textit{cat} components per logical state, leading to a minimal Hilbert space dimension scaling of $\sim 4d$, corresponding to a total spin quantum number $S = 2d-3/2$. This constraint expands to Hilbert space scaling of $\sim 12d$ when correcting all Pauli $X, Y$, and $Z$ errors. Increasing the code distance from $2t+1$ to $2(t+1)+1$ requries additional independent basis components to correct higher order errors, doubling the Hilbert space dimension at each step. The detailed construction with working examples follows.

We first present Z-error correcting codewords for qutrits and ququarts as working examples and provide a method for generalizing them to higher-dimensional qudits. In solid-state spin systems, when an external magnetic field is applied along the Z-axis to quantize individual $m_S, m_I$ levels, perturbations along the Z-axis often act as a dominant source of decoherence, i.e., the $T_2$ relaxation time limits the overall coherence. Therefore, error correction schemes that focus on Z-error can effectively enhance the performance of the system. In this case, logical qutrit codewords that counteract dephasing can be designed using a spin 9/2 system, as shown below,

\begin{equation}
\begin{split}
\ket{0_L} = \sqrt{\frac{10}{20}}( \ket*{-\frac{5}{2}} +  \ket*{+\frac{5}{2}}) \\
\ket{1_L} = \sqrt{\frac{6}{20}}( \ket*{-\frac{3}{2}} +  \ket*{+\frac{3}{2}}) + \sqrt{\frac{4}{20}}( \ket*{-\frac{7}{2}} +  \ket*{+\frac{7}{2}}) \\
\ket{2_L} = \sqrt{\frac{7}{20}}( \ket*{-\frac{1}{2}} +  \ket*{+\frac{1}{2}}) + \sqrt{\frac{3}{20}}( \ket*{-\frac{9}{2}} +  \ket*{+\frac{9}{2}}) .
\label{qutrit_Z_correction_codeword}
\end{split}
\end{equation}

\begin{figure}
\includegraphics[width=16cm]{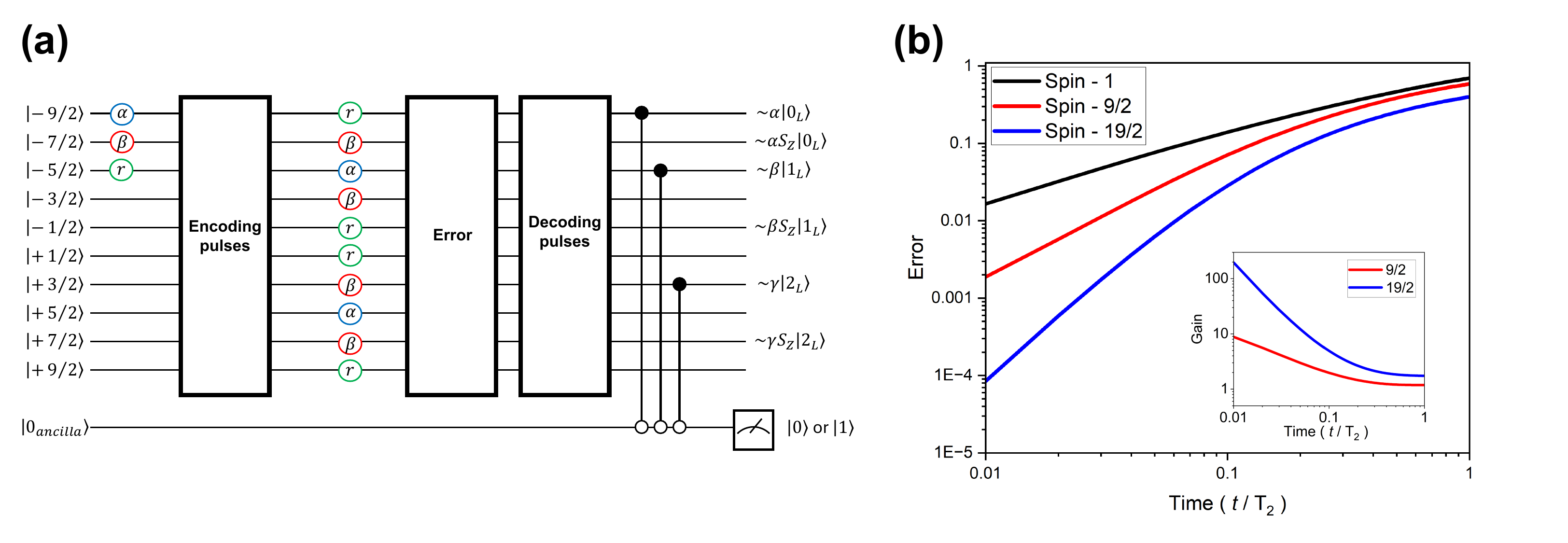}
\caption{(a) Brief schematic of qutrit Z-error correction code. (see main text for details) (b) The qutrit fidelity as a function of time with distance-1 (black), distance-3 (red), and distance-5 (blue) QEC code. Inset: Gain from fault-tolerant encoding, for distance-3 (red) and distance-5 (blue) encoding.
\label{qutrit_Z_error_fig}}
\end{figure}

Here, $\ket{m_I}$ corresponds to $Z$-basis eigenstates. The three code words presented above satisfy the qudit version of Knill-Laflamme (KL) criteria\cite{knill1997theory}, as shown below,

\begin{equation}
\begin{split}
\bra{i_L} E^{\dagger}_{a}E_{b} \ket{j_L} = 0 , (\mathrm{for~} i \neq j) \\
\bra{i_L} E^{\dagger}_{a}E_{b} \ket{i_L}  - \bra{j_L} E^{\dagger}_{a}E_{b} \ket{j_L} = 0 ,
\label{KL_criteria_eq}
\end{split}
\end{equation}

where $\ket{i_L},\ket{j_L}$ correspond to code words ($\ket{0_L}, \ket{1_L}, \ket{2_L}, \cdots$) and $E_a, E_b$ represent the target error operators. For this Z-error example, the relevant operators are $I$ and $S_Z$. In eq.\ref{qutrit_Z_correction_codeword}, all original codewords and error codewords are mutually orthogonal and form a pure error correction code. Therefore, both encoding and decoding can be implemented using simple pulse sequences. A brief schematic diagram of the overall process is shown in Fig. \ref{qutrit_Z_error_fig} (a), and the details of the encoding and decoding pulses are provided in Appendix~\ref{SI_chapter_A}. Without loss of generality, the original qutrit state can be prepared as $\ket{\psi} =  \alpha \ket*{-\frac{9}{2}} + \beta \ket*{-\frac{7}{2}} + \gamma \ket*{-\frac{5}{2}}$. This state can be encoded into the logical qutrit state, $\ket{\psi_{enc}} =  \alpha \ket*{0_L} + \beta \ket*{1_L} + \gamma \ket*{2_L}$, using the encoding pulses. During the storage time, Z error (i.e., phase error) may occur on the system, and the resulting state can be written as,

\begin{equation}
\begin{split}
E \ket{\psi_{enc}} =  \sqrt{1-\epsilon} (\alpha \ket{0_{L}}+ \beta \ket{1_{L}}+ \gamma \ket{2_{L}})      \\
+(i\epsilon_{Z})(\alpha S_Z\ket{0_{L}}+ \beta S_Z\ket{1_{L}}+ \gamma S_Z\ket{2_{L}}) \\
+(\frac{(i\epsilon_{Z})^2}{2!})(\alpha S_Z^2\ket{0_{L}}+ \beta S_Z^2\ket{1_{L}}+ \gamma S_Z^2\ket{2_{L}}) \\
+(\frac{(i\epsilon_{Z})^3}{3!})(\alpha S_Z^3\ket{0_{L}}+ \beta S_Z^3\ket{1_{L}}+ \gamma S_Z^3\ket{2_{L}}) \\
+\cdots \:, \\
\label{State_after_error}
\end{split}
\end{equation}

, where $\epsilon_Z$ corresponds to the error probability, i.e., $\frac{t}{T_{relax}}$. As in general quantum error correction protocols, we can assume that only the first and second terms in Eq~\ref{State_after_error} dominate the state for short times $t \ll T_{relax}$. Since all six terms ($\ket{0_{L}}, \ket{1_{L}}, \ket{2_{L}}, S_Z\ket{0_{L}}, S_Z\ket{1_{L}},$ and $S_Z\ket{2_{L}}$) are mutually orthogonal, the decoding pulses and conditional excitation of the ancilla qubit evolve the state to $\ket{\psi} =  A_0(\alpha \ket*{-\frac{9}{2}} + \beta \ket*{-\frac{5}{2}} + \gamma \ket*{+\frac{3}{2}})\ket{1_{ancilla}} + A_1(\alpha \ket*{-\frac{7}{2}} + \beta \ket*{-\frac{1}{2}} + \gamma \ket*{+\frac{7}{2}})\ket{0_{ancilla}}$. The projective measurement on the ancilla qubit then restores the original state, regardless of whether the outcome is 0 or 1.

This design can also be applied to ququarts or high-dimensional logical qudits. The encoding for a ququart can be done using the equation below, 

\begin{equation}
\begin{split}
\ket{0_L} = \sqrt{\frac{14}{28}}( \ket*{-\frac{7}{2}} +  \ket*{+\frac{7}{2}}) \\
\ket{1_L} = \sqrt{\frac{8}{28}}( \ket*{-\frac{5}{2}} +  \ket*{+\frac{5}{2}}) + \sqrt{\frac{6}{28}}( \ket*{-\frac{9}{2}} +  \ket*{+\frac{9}{2}}) \\
\ket{2_L} = \sqrt{\frac{9}{28}}( \ket*{-\frac{3}{2}} +  \ket*{+\frac{3}{2}}) + \sqrt{\frac{5}{28}}( \ket*{-\frac{11}{2}} +  \ket*{+\frac{11}{2}}) \\
\ket{3_L} = \sqrt{\frac{10}{28}}( \ket*{-\frac{1}{2}} +  \ket*{+\frac{1}{2}}) + \sqrt{\frac{4}{28}}( \ket*{-\frac{13}{2}} +  \ket*{+\frac{13}{2}}) .
\label{ququart_Z_correction_codeword}
\end{split}
\end{equation}

Using the same strategy, a $d$-dimensional logical qudit can be encoded with the code words ($\ket{0_L}, \ket{1_L}, \ket{2_L} ,\cdots, \ket{(d-1)_L} $) and coefficients as shown below,

\begin{equation}
\begin{split}
\ket{0_L} = \sqrt{\frac{1}{2}}( \ket*{-(S/2+1/4)} +  \ket*{+(S/2+1/4)} \\
\ket{i_L} = a_i( \ket*{-(S/2+1/4 -i )} +  \ket*{+(S/2+1/4 -i)}) + \\
 b_i( \ket*{-(S/2+1/4 +i)} + \ket*{+(S/2+1/4 + i)}), \\
\label{general_Z_correction_codeword}
\end{split}
\end{equation}

where $S = 2d -3/2$, and

\begin{equation}
\begin{split}
 a_i = \sqrt{\frac{2d-1+i}{8d-4}} \\
 b_i = \sqrt{\frac{2d-1-i}{8d-4}}. \\
\label{general_Z_correction_codeword_coefficients}
\end{split}
\end{equation}

These coefficients $a_i, b_i$ are defined to satisfy the KL criteria (see Appendix~\ref{SI_chapter_B}).

Another way to generalize the encoding scheme, beyond increasing the qudit dimension $d$, is to increase the distance of the error-correcting code. A higher-distance code allows the correction of more general or higher-order errors, for instance, up to second order errors using distance-5 codes. We provide a logical qutrit encoding that has distance-5, which can be realized using a spin S=19/2 system,

\begin{equation}
\begin{split}
\ket{0_L} = \sqrt{\frac{5}{16}}( \ket*{-\frac{5}{2}} +  \ket*{+\frac{5}{2}})+\sqrt{\frac{3}{16}}( \ket*{-\frac{15}{2}} +  \ket*{+\frac{15}{2}}) \\
\ket{1_L} = \sqrt{\frac{5423}{42400}}( \ket*{-\frac{1}{2}} +  \ket*{+\frac{1}{2}}) + \sqrt{\frac{7203}{42400}}( \ket*{-\frac{9}{2}} +  \ket*{+\frac{9}{2}}) \\
\sqrt{\frac{6517}{42400}}( \ket*{-\frac{11}{2}} +  \ket*{+\frac{11}{2}}) + \sqrt{\frac{2057}{42400}}( \ket*{-\frac{19}{2}} +  \ket*{+\frac{19}{2}}) \\
\ket{2_L} = \sqrt{\frac{3294}{22800}}( \ket*{-\frac{3}{2}} +  \ket*{+\frac{3}{2}}) + \sqrt{\frac{3749}{22800}}( \ket*{-\frac{7}{2}} +  \ket*{+\frac{7}{2}}) \\
 \sqrt{\frac{2771}{22800}}( \ket*{-\frac{13}{2}} +  \ket*{+\frac{13}{2}}) + \sqrt{\frac{1586}{22800}}( \ket*{-\frac{17}{2}} +  \ket*{+\frac{17}{2}}) .
\label{qutrit_Z_correction_codeword_second_order}
\end{split}
\end{equation}

The general form of distance-5 code and its coefficients for arbitrary $d$ are provided in Appendix~\ref{SI_chapter_B}, along with guidelines for designing codes with larger code-word distance $2t+1$.

In Fig.~\ref{qutrit_Z_error_fig} (b), we present the ideal performance of the logical qutrit encoding using spin 9/2 (Eq.~\ref{qutrit_Z_correction_codeword}) and 19/2 (Eq.~\ref{qutrit_Z_correction_codeword_second_order}) as a function of the ratio between the QEC scheme cycle time and $T_2$ relaxation time. The error ($ E=1-\mathcal{F}^2$) is obtained from natural dephasing using the Lindblad master equation. The inset of Fig.~\ref{qutrit_Z_error_fig} (b) shows the gain from the encoding, defined as $E_{\mathrm{uncorrected}} /E_{\mathrm{corrected}}$, providing evidence that the residual error decreases through larger qudit Hilbert spaces.

\section{$X, Y, Z$-error correction code for qutrit, and generalization}
\label{Chapter_XYZerror}

Next, we briefly discuss logical qudit encoding that can compensate for all Pauli $X, Y$, and $Z$ errors, enabling fully fault-tolerant circuits. The encoding can be designed based on a similar approach, but with additional spacing between code words to maintain orthogonality even in the presence of $X, Y$ errors. The simplest case of fault-tolerant qudit, i.e., qutrit, can be encoded using the following code word in a spin $S = 29/2$ system, 

\begin{equation}
\begin{split}
\ket{0_L} = \sqrt{\frac{10}{20}}( \ket*{-\frac{15}{2}} +  \ket*{+\frac{15}{2}}) \\
\ket{1_L} = \sqrt{\frac{6}{20}}( \ket*{-\frac{9}{2}} +  \ket*{+\frac{9}{2}}) + \sqrt{\frac{4}{20}}( \ket*{-\frac{21}{2}} +  \ket*{+\frac{21}{2}}) \\
\ket{2_L} = \sqrt{\frac{7}{20}}( \ket*{-\frac{3}{2}} +  \ket*{+\frac{3}{2}}) + \sqrt{\frac{3}{20}}( \ket*{-\frac{27}{2}} +  \ket*{+\frac{27}{2}}) .
\label{qutrit_XYZ_correction_codeword}
\end{split}
\end{equation}

This encoding shares the same set of coefficients as the $Z$-error correction encoding given in Eq.~\ref{qutrit_Z_correction_codeword}, but with a threefold larger Hilbert space dimension to provide additional spacing between code words. This encoding strategy can also be generalized to $d$-dimensional qudits, as shown in below,

\begin{equation}
\begin{split}
\ket{0_L} = \sqrt{\frac{1}{2}}( \ket*{-3 \cdot (S/2+1/4)} +  \ket*{+3 \cdot (S/2+1/4)} \\
\ket{i_L} = a_i( \ket*{-(3 \cdot (S/2+1/4)-3i)} +  \ket*{+(3 \cdot (S/2+1/4) - 3i) }) + \\
 b_i( \ket*{- (3 \cdot (S/2+1/4) + 3i)} + \ket*{+(3 \cdot (S/2+1/4) +3i) }), \\
\label{general_XYZ_correction_codeword}
\end{split}
\end{equation}

,with $S=6d-7/2$ and same set of $a_i, b_i$ coefficients in ~Eq.\ref{general_Z_correction_codeword_coefficients}.

While the designs above account for first-order errors, they can also be extended to second- and higher- order errors. This can be achieved by including additional Hilbert space dimensions and modifying the coefficients in the logical qudit definition. As a working example, logical qutrit encoding that can correct spin-operator errors up to second order can be implemented using a spin $S=99/2$, as shown below, 

\begin{equation}
\begin{split}
\ket{0_L} = \sqrt{\frac{5}{16}}( \ket*{-\frac{25}{2}} +  \ket*{+\frac{25}{2}})+\sqrt{\frac{3}{16}}( \ket*{-\frac{75}{2}} +  \ket*{+\frac{75}{2}}) \\
\ket{1_L} = \sqrt{\frac{5423}{42400}}( \ket*{-\frac{5}{2}} +  \ket*{+\frac{5}{2}}) + \sqrt{\frac{7203}{42400}}( \ket*{-\frac{45}{2}} +  \ket*{+\frac{45}{2}}) \\
\sqrt{\frac{6517}{42400}}( \ket*{-\frac{55}{2}} +  \ket*{+\frac{55}{2}}) + \sqrt{\frac{2057}{42400}}( \ket*{-\frac{95}{2}} +  \ket*{+\frac{95}{2}}) \\
\ket{2_L} = \sqrt{\frac{3294}{22800}}( \ket*{-\frac{15}{2}} +  \ket*{+\frac{15}{2}}) + \sqrt{\frac{3749}{22800}}( \ket*{-\frac{35}{2}} +  \ket*{+\frac{35}{2}}) \\
 \sqrt{\frac{2771}{22800}}( \ket*{-\frac{65}{2}} +  \ket*{+\frac{65}{2}}) + \sqrt{\frac{1586}{22800}}( \ket*{-\frac{85}{2}} +  \ket*{+\frac{85}{2}}) .
\label{qutrit_XYZ_correction_codeword_second_order}
\end{split}
\end{equation}

Similar to first-order error correction codeword, this design also shares the same coefficients as second-order Z-error codewords (i.e., Eq.~\ref{qutrit_Z_correction_codeword_second_order}), but with a fivefold increase in Hilbert space dimension and spacing between basis states. In general, we find that once the distance $2t+1$ Z-error correction codewords are specified, code words for all X/Y/Z error corrections with the same distance can be directly obtained by increasing all basis indices by a factor of $2t+1$ in the Z-basis. (See Appendix~\ref{SI_chapter_B})

Depending on the quantum platform, entangled spin qudits can also be employed to construct highly coherent logical qudits. Here, we provide several examples, such as a three-spin-9/2 qudit code that can encodes a fault-tolerant qutrit,

\begin{equation}
\begin{split}
\ket{0_L} = \sqrt{\frac{10}{20}}( \ket*{-\frac{5}{2}}_{A,B,C} +  \ket*{+\frac{5}{2}}_{A,B,C}) \\
\ket{1_L} = \sqrt{\frac{6}{20}}( \ket*{-\frac{3}{2}}_{A,B,C} +  \ket*{+\frac{3}{2}}_{A,B,C}) + \sqrt{\frac{4}{20}}( \ket*{-\frac{7}{2}}_{A,B,C} +  \ket*{+\frac{7}{2}}_{A,B,C}) \\
\ket{2_L} = \sqrt{\frac{7}{20}}( \ket*{-\frac{1}{2}}_{A,B,C} +  \ket*{+\frac{1}{2}}_{A,B,C}) + \sqrt{\frac{3}{20}}( \ket*{-\frac{9}{2}}_{A,B,C} +  \ket*{+\frac{9}{2}}_{A,B,C}) .
\label{three_qutrit_Z_correction_codeword}
\end{split}
\end{equation}

where A, B, and C denote the three qudits, and $\ket{m_I}_{A,B,C}$ corresponds to the entangled product state $\ket{m_I}_A \otimes \ket{m_I}_B \otimes \ket{m_I}_C$. The multi-qudit encoding and decoding pulse sequences can be designed in the same manner as proposed in Ref.~\cite{lim2025designing}. The above logical qutrit codewords with three $S=9/2$ systems have the advantage of being implementable using nuclear spins in solid-states platforms, which are among the most coherent quantum systems.

This multi-spin qudit code can also be extended to second-order correction, as shown in the encoding with five 19/2 spins below,

\begin{equation}
\begin{split}
\ket{0_L} = \sqrt{\frac{5}{16}}( \ket*{-\frac{5}{2}}_{A,B,C,D,E} +  \ket*{+\frac{5}{2}}_{A,B,C,D,E} )+\sqrt{\frac{3}{16}}( \ket*{-\frac{15}{2}}_{A,B,C,D,E}  +  \ket*{+\frac{15}{2}}_{A,B,C,D,E} ) \\
\ket{1_L} = \sqrt{\frac{5423}{42400}}( \ket*{-\frac{1}{2}}_{A,B,C,D,E}  +  \ket*{+\frac{1}{2}}_{A,B,C,D,E} ) + \sqrt{\frac{7203}{42400}}( \ket*{-\frac{9}{2}}_{A,B,C,D,E}  +  \ket*{+\frac{9}{2}}_{A,B,C,D,E} ) \\
\sqrt{\frac{6517}{42400}}( \ket*{-\frac{11}{2}}_{A,B,C,D,E}  +  \ket*{+\frac{11}{2}}_{A,B,C,D,E} ) +  \sqrt{\frac{2057}{42400}}( \ket*{-\frac{19}{2}}_{A,B,C,D,E}  +  \ket*{+\frac{19}{2}}_{A,B,C,D,E} ) \\
\ket{2_L} = \sqrt{\frac{3294}{22800}}( \ket*{-\frac{3}{2}}_{A,B,C,D,E}  +  \ket*{+\frac{3}{2}}_{A,B,C,D,E} ) +  \sqrt{\frac{3749}{22800}}( \ket*{-\frac{7}{2}}_{A,B,C,D,E}  +  \ket*{+\frac{7}{2}}_{A,B,C,D,E} ) \\
\sqrt{\frac{2771}{22800}}( \ket*{-\frac{13}{2}}_{A,B,C,D,E}  +  \ket*{+\frac{13}{2}}_{A,B,C,D,E} ) +\sqrt{\frac{1586}{22800}}( \ket*{-\frac{17}{2}}_{A,B,C,D,E}  +  \ket*{+\frac{17}{2}}_{A,B,C,D,E} ) .
\label{qutrit_Z_correction_codeword_second_order_five}
\end{split}
\end{equation}

where the notation $\ket{m_I}_{A,B,C,D,E}$ denotes the tensor product state of five spin qudits, as in the previous case. 

\section{Discussions on dimensions, practical implementations, and logical qudit operations}
\label{Chapter_Discussion}

Building on the encoding schemes introduced in Sections~\ref{Chapter_Zerror} and~\ref{Chapter_XYZerror}, this chapter discusses the required Hilbert-space dimensionality, practical considerations for physical implementation, and logical qudit gate operations. In addition, we identify the minimum single-qudit gate fidelities required for fault-tolerant encoding to yield a net advantage. 

The conventional definition of a code distance $2t+1$ in quantum error correction is often interpreted as the ability to correct $t$ simultaneous and independent errors in logical qubit encodings. This definition relies on two characteristic properties of qubits: first, that the square of qubit Pauli operators is the identity, and second, that logical qubit encoding requires multiple physical qubits. These assumptions, however, are often not valid in qudit systems. A logical unit can be encoded within a single qudit, and the square of error operators is not necessarily the identity. 

The first and most well-known example illustrating this distinction is the GKP code\cite{gottesman2001encoding}. The GKP code requires a bosonic Hilbert space\cite{pirandola2008minimal, michael2016new}, as implied by the definition of the generalized $d$-dimensional Pauli operators,

\begin{equation}
\begin{split}
X\ket{j} = \ket{j+1 ~(\mathrm{mod~} d)},  \\
Z\ket{j} = w^j\ket{j} ,
\label{Generalized_Pauli}
\end{split}
\end{equation}

where $d$ denotes the qudit dimension embedded as a periodic subset of an infinite-dimensional Hilbert space, and $w$ is a phase factor. In this framework, higher-order physical errors in the system can be expressed as combinations of generalized $X$ and $Z$ operators up to order $t$, and are therefore correctable using a codeword with distance $2t+1$. 

While this mathematical structure is also well-defined for finite-dimensional qudits, its physical implementation requires translational invariance (i.e., periodic boundary conditions) in Hilbert space. In particular, unknown physical errors in finite-dimensional systems, such as spin qudits, do not generate cyclic shift operations across the Hilbert space boundary. This issue is mitigated in the case of harmonic oscillators by encoding into infinite-dimensional eigenstates with finite squeezing, which effectively preserves translational symmetry. This mechanism, however, is not available in intrinsically finite-dimensional systems. 

Indeed, this consideration motivated the alternative encodings described in the previous chapters. A similar perspective on codeword distance can be applied to the spin Hamiltonian of the system. In solid-state spin systems, the dominant Hamiltonian terms governing the system can often be written as,

\begin{equation}
H = g \mu_B B \cdot S + g_I \mu_B B \cdot I + S \cdot A \cdot I + S \cdot D \cdot S + I \cdot Q \cdot I + \mathrm{(higher~order~terms)} 
\label{Si_Sb_H}
\end{equation}    

where $I, S$ represent the nuclear and electronic spin operators, respectively. From left to right, the terms describe the electronic and nuclear spin Zeeman splittings, the hyperfine interaction, the second-order crystal field splitting, and the nuclear quadrupole interaction. While additional higher-order terms (such as crystal field splitting represented in terms of Stevens operators) may arise from the crystalline symmetry and environmental effects, these operators can generally be decomposed into combinations of $I$ and $S$ operators. These terms can act as qudit control operations under engineered perturbations are applied\cite{fernandez2024navigating}, and as error sources under unknown perturbations\cite{lim2025designing}. Higher-order errors generated by $I,S$ operators up to order $\sim t$ can therefore be corrected using a logical qudit encoding with distance of $2t+1$.


\begin{figure}
\includegraphics[width=16cm]{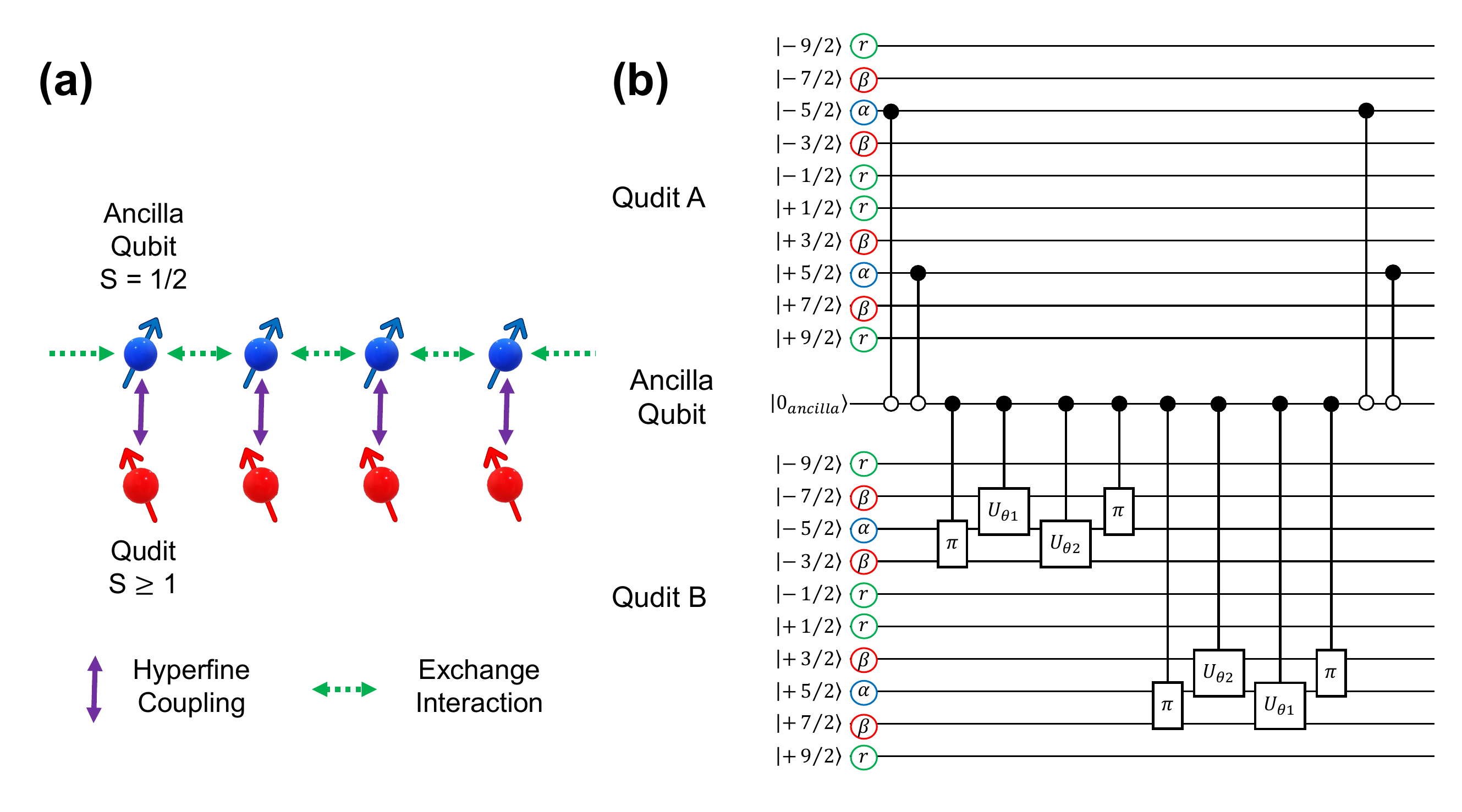}
\caption{(a) Schematic illustration of a scalable multi-qudit architecture (b) Example of a logical two-qudit gate (CNOT) implemented via an ancilla qubit as a mediator. Unitary rotations $U_{\theta i}$ are around $y$ axis, with $cos(\theta_i) = \sqrt{2/5}, \sqrt{3/5}$ for $i = 1,2$, respectively.(see main text for details)
\label{qudit_scheme_fig}}
\end{figure}

Next, we discuss the resources required for practical quantum computation based on these logical qudits. These include the number of physical units (qubits or qudits), the number of elementary gate operations required to decode and correct a single logical qudit, and the implementation of logical qudit gates. As illustrated in Fig.~\ref{qudit_scheme_fig} (a), one possible architecture consists of each qudit being coupled to a nearest ancilla qubit. This ancilla serves both as a resource for error detection and as a mediator of inter-qudit interactions. 

Under this connectivity condition, a distance-3 logical qudit code requires on the order of $\sim d^2$ elementary operations for decoding, error correction, and logical single- and two-qudit gates. The decoding and correction procedures are discussed in Chapter II and Appendix.~A. Logical single-qudit operations can be implemented as unitary transformations between code words. Since all original and error codewords are mutually orthogonal in this encoding, the corresponding unitary transformations can be explicitly constructed and implemented with at most $\sim d^2$ elementary operations. Logical two-qudit gates can be implemented via an ancilla qubit, as depicted in Fig.~\ref{qudit_scheme_fig} (b). The sequence shown corresponds to a controlled-NOT operation from $\ket{0_L}_A$ to $\ket{0_L}_B \leftrightarrow \ket{1_L}_B$. Importantly, this operation does not depend on the presence of $Z$ error on control qudit A, allowing error correction to be performed either before or after the two-qudit operation. For the case of $X, Y$, and $Z$ error correction encoding of spin 29/2 system (Eq.~\ref{qutrit_XYZ_correction_codeword}), the same operation can be achieved by adding nearest-level controlled-NOT operations from qudit A to the ancilla qubit. Specifically, these correspond to CNOT operations conditioned on the levels $\ket{-17/2}, \ket{-15/2}, \ket{-13/2}, \ket{+13/2}, \ket{+15/2}$, and $\ket{+17/2}$ of qudit A. Other controlled operations, such as $C_iNOT_{j,k}$ with different indices $i,j,k$, can be implemented in the same manner within $\sim d^2$ elementary operations. Since a universal gate set for the qudits consists of arbitrary single-qudit gates between adjacent levels and two-qudit entangling operations such as $C_iNOT_{j,k}$ between arbitrary levels\cite{brennen2005criteria, wang2020qudits}, all logical operations can be constructed with polynomial complexity scaling in $d$. 

This type of connectivity between spin qudits can be experimentally implemented in state-of-the-art platforms, including hybrid nuclear-electronic spin systems, where the system Hamiltonian can be described by an extension of ~Eq.\ref{Si_Sb_H}, with an additional exchange interaction between electron spins, $JS_1 \cdot S_2$. Controlled transitions between nuclear spin qudits and hyperfine-coupled electronic spin qubits have been experimentally demonstrated in paramagnetic impurities in semiconductor systems \cite{lim2025demonstrating, yu2025schrodinger} as well as in molecular spin qudits\cite{rubin2025implementation, chicco2023proof}. For inter-qudit connectivity, electron spins coupled via exchange interactions can serve as mediators of entanglement between two nuclear spins, as originally proposed theoretically in the Kane model \cite{kane1998silicon} and realized experimentally in semiconductor quantum dot spin systems\cite{stemp2025scalable}. While such implementations require multiple electron spins to mediate interactions, this additional overhead scales only linearly with the number of elementary operations required for logical two-qudit gates. Regarding the addressing of the individual spin sublevels, while first-order Zeeman terms in ~Eq.\ref{Si_Sb_H} lead to uniform energy level splittings, additional perturbative terms - such as hyperfine interactions, crystal field effects, nuclear quadrupole interactions - provide spectral addressability of the sublevels\cite{lim2025demonstrating, fernandez2024navigating}. Fault-tolerant encoding of a qutrit can be implemented in several nuclear-spin environments, such as Bi dopants in Si systems, and higher-dimensional qudits require access to larger Hilbert spaces. In such cases, the qudit space can be realized through the chemical design of giant molecular magnets\cite{leuenberger2001quantum, takahashi2009coherent, murugesu2008large}, or by using several coupled nuclear spin systems \cite{rahmer2005w, liu2018qubit}. Additional candidate platforms using similar spin Hamiltonians include trapped and laser-cooled ions or molecules (which possess additional total angular momentum contributions from motional degrees of freedom) \cite{boguslawski2023raman, yu2022magneto}, and photonic systems utilizing orbital angular momentum states\cite{yao2011orbital, kim2024qudit}.

The features discussed above indicate that logical qudit operations can be implemented with polynomial complexity in the qudit dimension $d$, while maintaining similar or lower overhead compared to conventional mapping from logical qubit encodings. Importantly, this approach retains the advantage of requiring fewer physical units to be controlled. This resource-efficiency can be further quantified by comparing the total Hilbert space dimension required for logical qudit construction, as discussed Fig.~\ref{SI_chapter_C_fig} and Appendix sections \ref{SI_chapter_C}. The total Hilbert space dimension provides an upper bound on the physical state space accessed during decoding and is directly relevant to experimental complexity of physical implementation. Regarding gate complexity in information processing, there has been extensive work on decomposing of qudit-based gates into qubit-based gates\cite{kiktenko2025colloquium}. However, several recent studies suggest that qudit-to-qubit mapping typically requires extra resources, such as extra entangling gates in Hamiltonian exponentiation\cite{fischer2023universal, chizzini2024qudit}, or increased circuit depth in Grover's algorithm implementations\cite{rambow2109reduction}. These  results further support the conclusion that the direct use of logical qudits can offer practical advantages in experimental algorithm implementation.

\begin{figure}
\includegraphics[width=8cm]{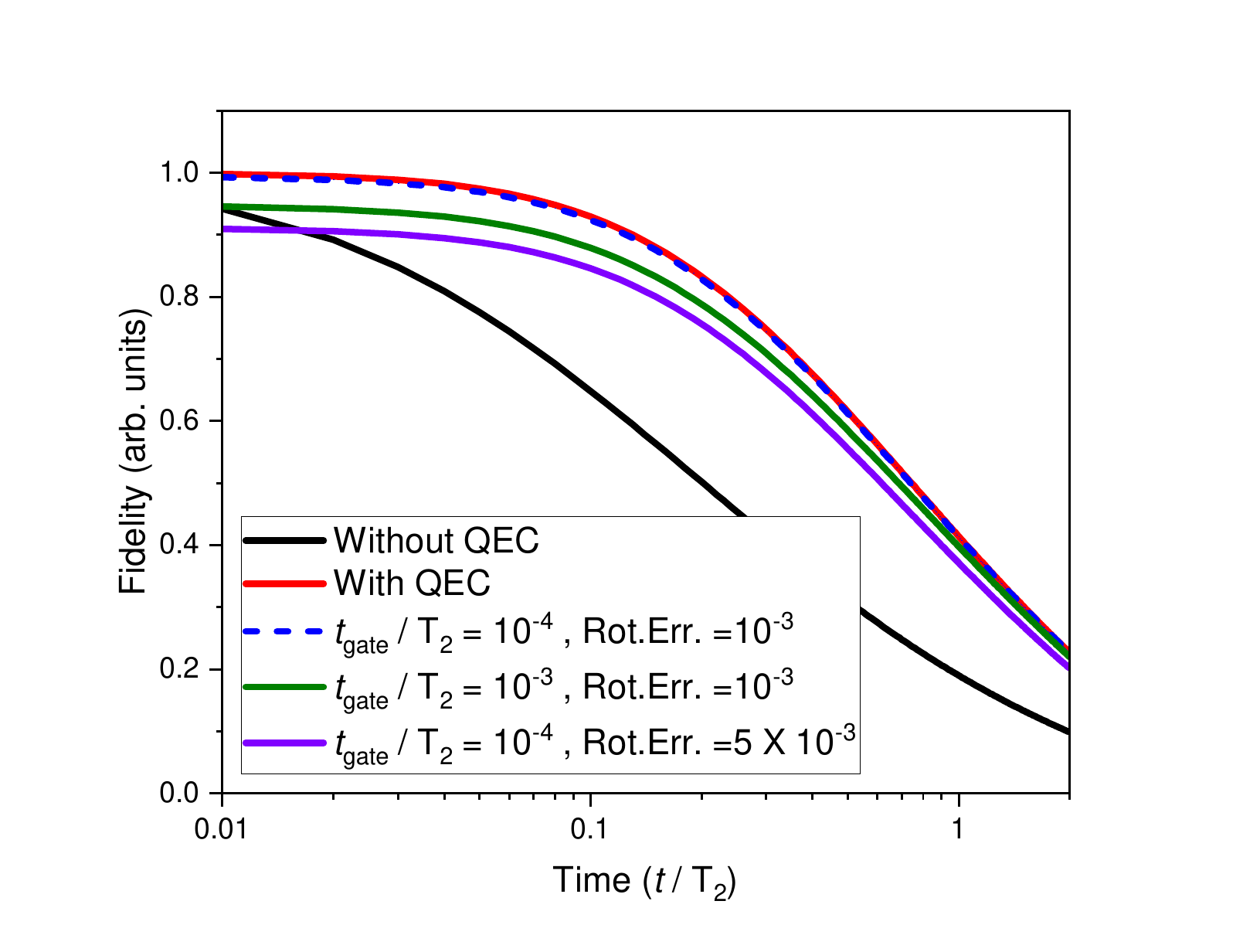}
\caption{ Qutrit fidelity as a function of time without (black) and with (red) error correction. The effects of decoherence during the decoding sequence and imperfect gate rotation angle errors are also simulated, shown as green and purple solid lines, and a blue dotted line (see main text for details). 
\label{qutrit_requried_gate_condition}}
\end{figure}

In Fig.~\ref{qutrit_requried_gate_condition}, we present a numerical estimate of the overall fidelity of the logical qutrit code defined in Eq.~\ref{qutrit_Z_correction_codeword}. The time evolution of a superposed qutrit state ($\ket{\psi} = \frac{1}{\sqrt{3}}( \ket{0_L} + \ket{1_L} + \ket{2_L} ) $), encoded in a nuclear spin 9/2 qudit coupled to an electron spin-1/2 ancilla qubit, is simulated with a typical spin-spin relaxation time, $T_2$. While more specific microscopic models of spin decoherence exist for various experimental environments\cite{onizhuk2025colloquium}, we model decoherence using a Lindblad master equation, assuming fluctuations of the longitudinal magnetic field $B_Z$, as the dominant decoherence source. This model corresponds to the conventional case of well-isolated spin qudit memories in the solid-state. It is typically observed in the decay of electronic and nuclear spins of defects in semiconductors\cite{lim2025demonstrating}, or in a molecualr spin qudits\cite{chiesa2020molecular}, which exhibit millisecond-scale coherence for nuclear spins and tens of microseconds for electronic spins in a typical X-band ($\sim 0.3$ Tesla) EPR setup.  

A comparison of the fidelities with (red) and without (black) the error correction protocol confirms the effectiveness of the encoding in the regime $ t \ll T_2$. We further test a model that accounts for single-qudit gate fidelity and decoherence during the encoding and decoding process, in order to identify practical conditions to achieve a net advantage with the error correction protocol. In the simulation, we have assumed that only the $\delta m_I = 1$ transition is allowed in the system ( which could be further improved if the system allows $\delta m_I = 2$ transitions), and gate infidelity is defined as rotation angle and phase error of these elementary gates. As shown by comparisons across different gate fidelities and single gate duration parameters - such as single gate fidelity of $99.5 \%$ (purple) and single gate duration $t_{gate} / T_2 = 10^{-3}$ (green)-- we find that the optimal protocol requires a single gate fidelity above $99.9 \%$, and single gate duration shorter than $10^{-4}$ times $T_2$ (blue dotted). In conventional EPR and NMR setups, single gate duration condition is already achieved\cite{lim2025demonstrating}, and state-of-the-art spin qubit devices report single gate fidelities up to two to three nines\cite{yoneda2018quantum, noiri2022fast}. Therefore we believe that logical qudit encoding is readily implementable, and can offer an advantage over conventional qubit-to-qudit mapping schemes.

\begin{acknowledgments}
We thank Arzhang Ardavan for helpful discussions. This project was supported by the European Union's Horizon 2020 research and innovation programme under grant agreements 862893 (FATMOLS) and 863098 (SPRING), the National Research Foundation of Korea (NRF) under Grant No. RS-2023-00256050.
\end{acknowledgments}


\bibliography{Fault_to}



\appendix

\setcounter{figure}{0}
\renewcommand{\figurename}{Fig.}
\renewcommand{\thefigure}{A\arabic{figure}}

\section{Pulse sequences for encoding and decoding logical qudits }
\label{SI_chapter_A}

\begin{figure}
\includegraphics[width=16cm]{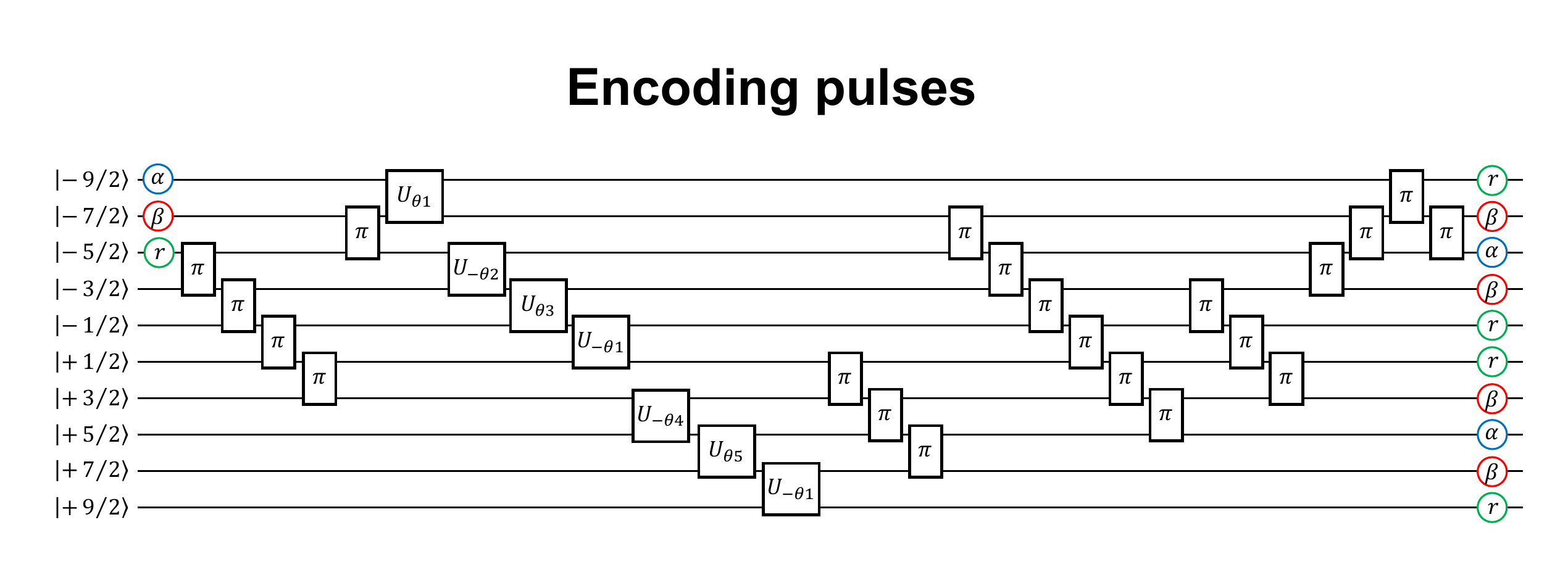}
\caption{ Gates used in the encoding pulses sequence of the qutrit Z-error correction code. Unitary rotations $U_{\theta i}$ are around $y$ axis, with $cos(\theta_i) = \sqrt{1/2}, \sqrt{3/10}, \sqrt{3/7}, \sqrt{7/20}, \sqrt{7/13}$ for $i = 1,2,3,4,5$, respectively.
\label{ENC_pulse}}
\end{figure}

Fig.~\ref{ENC_pulse} shows the complete encoding pulse sequence for the qutrit Z-error correction code. Without loss of generality, the initial state can be written as  $\ket{\psi} =  \alpha \ket*{-\frac{9}{2}} + \beta \ket*{-\frac{7}{2}} + \gamma \ket*{-\frac{5}{2}}$.

Through the encoding sequence, the general starting state is transformed into $\ket{\psi} =  \alpha \ket*{0_L} + \beta \ket*{1_L} + \gamma \ket*{2_L}$.

\begin{figure}
\includegraphics[width=16cm]{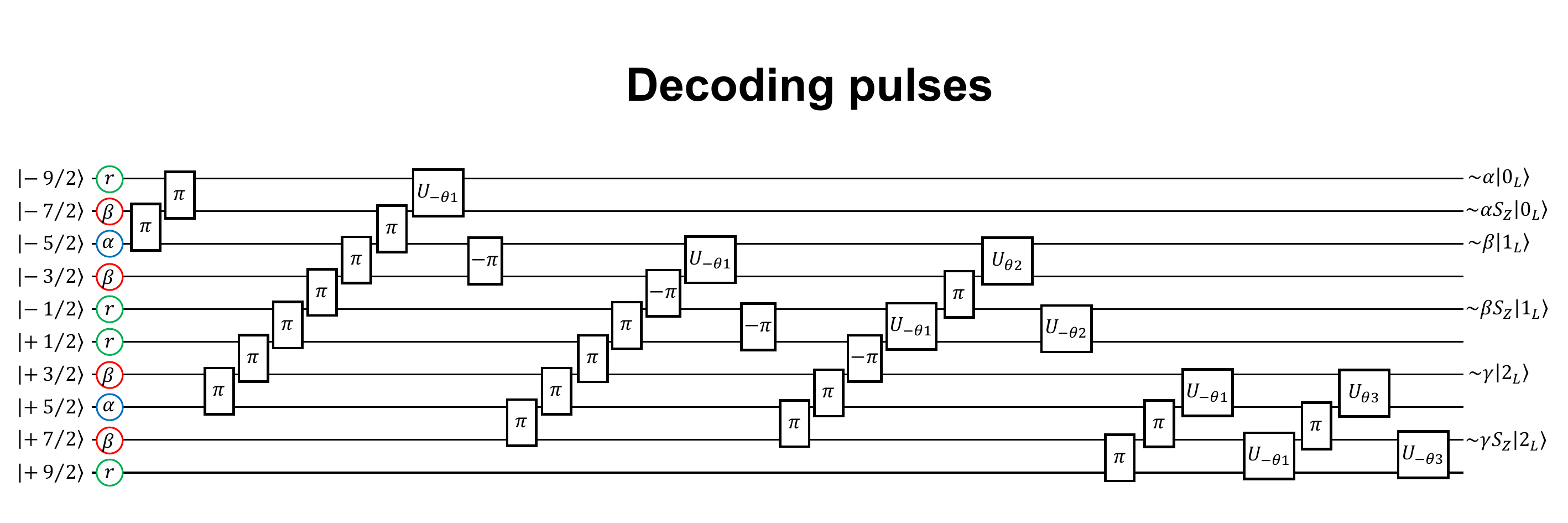}
\caption{ Gates used in the decoding pulse sequence of the qutrit Z-error correction code. Unitary rotations $U_{\theta i}$ are around $y$ axis, with $cos(\theta_i) = \sqrt{1/2}, \sqrt{2/5}, \sqrt{3/10}$ for $i = 1,2,3$, respectively.
\label{DEC_pulse}}
\end{figure}

Fig.~\ref{DEC_pulse} shows the complete decoding pulse sequence. The state after Z-error on the encoded state can be written as, 

\begin{equation}
\begin{split}
E \ket{\psi_{enc}} =  \sqrt{1-\epsilon} (\alpha \ket{0_{L}}+ \beta \ket{1_{L}}+ \gamma \ket{2_{L}})      \\
+(i\epsilon_{Z})(\alpha S_Z\ket{0_{L}}+ \beta S_Z\ket{1_{L}}+ \gamma S_Z\ket{2_{L}}) \\
+\mathrm{(higher~order~terms)} \:. \\
\label{State_after_error_SI}
\end{split}
\end{equation}

After applying the decoding pulses, the state is transformed into $\ket{\psi} =  A_0(\alpha \ket*{-\frac{9}{2}} + \beta \ket*{-\frac{5}{2}} + \gamma \ket*{+\frac{3}{2}}) + A_1(\alpha \ket*{-\frac{7}{2}} + \beta \ket*{-\frac{1}{2}} + \gamma \ket*{+\frac{7}{2}})$.

\section{The coefficients for logical qudit encoding code words}
\label{SI_chapter_B}

In this section, we present the detailed derivation of the logical qudit codeword construction, including formalism of the $Z$-basis combinations and their coefficients, and the generalization to arbitrary qudit dimension $d$ and codeword distance $2t+1$. As a working example, we consider the first-order (i.e., distance-3) Z-error correction code, for which each logical qudit state can be written as,

\begin{equation}
\begin{split}
\ket{0_L} = \sqrt{\frac{1}{2}}( \ket*{-(S/2+1/4)} +  \ket*{+(S/2+1/4)}) \\
\ket{i_L} = a_i( \ket*{-(S/2+1/4 -i )} +  \ket*{+(S/2+1/4 -i)}) + \\
 b_i( \ket*{-(S/2+1/4 +i)} + \ket*{+(S/2+1/4 + i)}), \\
\label{general_Z_correction_codeword_again}
\end{split}
\end{equation}

where $S = 2d -3/2$ and $i = 1,2, \cdots , d-1$. The coefficients $a_i$ and $b_i$ must be carefully chosen to satisfy the KL criteria defined in Eq.\ref{KL_criteria_eq}. For the first-order Z-error case, the nontrivial conditions in the KL criteria can be summarized as the following three conditions: 1) $\bra{i_L}\ket{i_L}=1$ (Normalization), 2) $\bra{i_L}S_Z\ket{i_L}=\bra{j_L}S_Z\ket{j_L}$ , 3) $\bra{i_L}S_Z^2\ket{i_L}=\bra{j_L}S_Z^2\ket{j_L}$, for $i,j = 0,1,2$. It follows straightforwardly that the coefficients $b_i$ are determined from $a_i$ using normalization condition. The first-order projection condtions are automatically satisfied by symmetry. The second-order projection conditions constrain $a_i$ coefficients through the relation of $\bra{0_L}S_Z^2\ket{0_L}=\bra{i_L}S_Z^2\ket{i_L}$. Therefore, choosing the total angular momentum $S=2d-3/2$ yields a valid solution, as shown in Eq.~\ref{general_Z_correction_codeword_coefficients}.

To construct the distance-5 code, we double the number of $S_Z$-basis states by duplicating the distribution used in the distance-3 code and placing it in the higher-$S_Z$ subspace. For example, the distance-5 code for a qutrit retains the codeword distribution from $\ket{1/2} \cdots \ket{9/2}$ and adds a similar set spanning $\ket{11/2} \cdots \ket{19/2}$. 

In this distance-5 case, the codewords must satisfy the KL conditions $\bra{i_L}\ket{i_L}=1$, and $\bra{i_L}S_Z^n\ket{i_L}=\bra{j_L}S_Z^n\ket{j_L}$, for $n=1, 2, 3, 4$. The general form of the distance-5 code for a $d$-dimensional qudit can be written as,

\begin{equation}
\begin{split}
\ket{0_L} = \sqrt{\frac{5}{16}}( \ket*{-(S/4+1/8)} +  \ket*{+(S/4+1/8)}) \\
+ \sqrt{\frac{3}{16}}( \ket*{-(3S/4+3/8)} +  \ket*{+(3S/4+3/8)}) \\
\ket{i_L} = a_i( \ket*{-(2d - 1/2 - i )} +  \ket*{+(2d - 1/2 - i )})  \\
 + b_i( \ket*{-(2d - 3/2 + i )} + \ket*{+(2d -3/2 + i )}) \\
 + c_i( \ket*{-(i -1/2)} + \ket*{+(i -1/2)}), \\
 + d_i( \ket*{-(4d-3/2 - i)} + \ket*{+(4d-3/2 - i)}), \\
\label{general_Z_second_correction_codeword}
\end{split}
\end{equation}

where $S=4d-5/2$, and $i = 1,2, \cdots , d-1$. The coefficients satisfying the above KL criteria are given by,

\begin{equation}
\begin{split}
 a_i = \sqrt{\frac{8d+2i-5}{32d-16}} \cdot \sqrt{\frac{66d^3+2d^2i-100d^2-14di^2+12di+47d+2i^3+4i^2-5i-7}{96d^3+32d^2i-160d^2-16di^2-16di+84d+8i^2-14}} \\
 b_i = \sqrt{\frac{8d-2i-3}{32d-16}} \cdot \sqrt{\frac{66d^3+2d^2i-100d^2-14di^2+12di+47d+2i^3+4i^2-5i-7}{96d^3+32d^2i-160d^2-16di^2-16di+84d+8i^2-14}} \\
 c_i = \sqrt{\frac{12d-2i-5}{32d-16}} \cdot \sqrt{\frac{36d^3+30d^2i-60d^2-2di^2-28di+37d-2i^3+4i^2+5i-7}{96d^3+32d^2i-160d^2-16di^2-16di+84d+8i^2-14}}\\
 d_i = \sqrt{\frac{4d+2i-3}{32d-16}} \cdot \sqrt{\frac{36d^3+30d^2i-60d^2-2di^2-28di+37d-2i^3+4i^2+5i-7}{96d^3+32d^2i-160d^2-16di^2-16di+84d+8i^2-14}}. \\
\label{general_Z_correction_codeword_coefficients}
\end{split}
\end{equation}

Although there are five sets of conditions in the KL criteria, the odd-order error projection terms ($\bra{i_L}S_Z\ket{i_L}=\bra{j_L}S_Z\ket{j_L}$, $\bra{i_L}S_Z^3\ket{i_L}=\bra{j_L}S_Z^3\ket{j_L}$) are automatically satisfied by symmetry. Therefore, it is sufficient to verify only the remaining even-order conditions, $\bra{i_L}\ket{i_L}=1$, $\bra{i_L}S_Z^2\ket{i_L}=\bra{j_L}S_Z^2\ket{j_L}$, and $\bra{i_L}S_Z^4\ket{i_L}=\bra{j_L}S_Z^4\ket{j_L}$. As a result, the above codeword design with four free parameters satisfies all KL criteria. 

It is worth noting that the distance-5 code for a qutrit, as well as for general $d$-dimensional qudits, is not unique. As a simple example, we present an alternative set of coefficients,

\begin{equation}
\begin{split}
\ket{0_L} = \sqrt{\frac{3}{10}}( \ket*{-\frac{5}{2}} +  \ket*{+\frac{5}{2}})+\sqrt{\frac{1}{5}}( \ket*{-\frac{15}{2}} +  \ket*{+\frac{15}{2}}) \\
\ket{1_L} = \sqrt{\frac{1152}{9225}}( \ket*{-\frac{1}{2}} +  \ket*{+\frac{1}{2}}) + \sqrt{\frac{133}{1025}}( \ket*{-\frac{9}{2}} +  \ket*{+\frac{9}{2}}) \\
\sqrt{\frac{399}{2050}}( \ket*{-\frac{11}{2}} +  \ket*{+\frac{11}{2}}) + \sqrt{\frac{468}{9225}}( \ket*{-\frac{19}{2}} +  \ket*{+\frac{19}{2}}) \\
\ket{2_L} = \sqrt{\frac{1081}{7700}}( \ket*{-\frac{3}{2}} +  \ket*{+\frac{3}{2}}) + \sqrt{\frac{252}{1650}}( \ket*{-\frac{7}{2}} +  \ket*{+\frac{7}{2}}) \\
 \sqrt{\frac{441}{3300}}( \ket*{-\frac{13}{2}} +  \ket*{+\frac{13}{2}}) + \sqrt{\frac{282}{3850}}( \ket*{-\frac{17}{2}} +  \ket*{+\frac{17}{2}}) .
\label{qutrit_Z_correction_codeword_second_order_other}
\end{split}
\end{equation}

For general code-word distance of $2t+1$ with $t \ge 3$, similar expansion -- which doubles the encoding $S_Z$-basis in the same manner as the distance-3 to distance-5 construction -- provides sufficient Hilbert space and degrees of freedom to satisfy the KL criteria, i.e., $\bra{i_L}S_Z^n\ket{i_L}=\bra{j_L}S_Z^n\ket{j_L}$ for $n=0,1,2 \cdots 2t$, by appropriately tailoring the coefficients. As in the above case, for odd $t$, the condition $\bra{i_L}S_Z^t\ket{i_L}=\bra{j_L}S_Z^t\ket{j_L}$ is trivially satisfied from symmetry. Therefore, the remaining nontrivial KL conditions correspond to even orders of $t$. Based on this iterative doubling of the codeword structure, an arbitrary $\ket{i_L}$ consists of a linear combination of $2^t$ Z-basis states, with the same number of independent coefficients. Pairing the coefficients $t$ times across $2^t$ basis states provide sufficient degrees of freedom to construct arbitrary $\ket{i_L}$ satisfying $\bra{i_L}S_Z^n\ket{i_L}=\bra{0_L}S_Z^n\ket{0_L}$ for $n= 0, 2, \cdots 2t$.  We emphasize that this systematic procedure provides a sufficient condition for constructing valid codewords. Further investigation is required to identify the mathematically minimal Hilbert space dimension that enables fault-tolerant encoding.

We presented a general formalism for extending the $Z$-error-correcting code to full $X/Y/Z$ error correction. This is achieved by increasing the basis dimension by a factor of $2t+1$ (i.e., a threefold increase from 9/2 code to the 29/2 code, and a fivefold increase from 19/2 code to 99/2), or by increasing the number of qudits by a factor of $2t+1$ (i.e., three-9/2 qudits code or five-19/2 qudits code). In both approaches, the same set of coefficients used for the $Z$-error correction remains valid for constructing the corresponding distance-$2t+1$ $X/Y/Z$ error-correcting code. 

As a working example, we briefly describe why the code-words in Eq.~\ref{qutrit_Z_correction_codeword} can also be extended to correct X/Y/Z errors. The Eq.~\ref{qutrit_Z_correction_codeword} satisfies the conditions $\bra{i_L}\ket{i_L}=1$ , $\bra{i_L}S_Z\ket{i_L}=\bra{j_L}S_Z\ket{j_L}$ , $\bra{i_L}S_Z^2\ket{i_L}=\bra{j_L}S_Z^2\ket{j_L}$, for $i,j = 0,1,2$. Therefore, the threefold expanded codewords in Eq.\ref{qutrit_XYZ_correction_codeword} automatically satisfy the above three conditions, since the expansion corresponds to a linear scaling of the $S_Z$-basis. The orthogonality conditions in the KL criteria ( $\bra{i_L}E_aE_B\ket{j_L}=0$) are also fulfilled due to the $2t+1$-fold spacing between the basis states, for any $E_a, E_b$ corresponding to X, Y, or Z operator. Thus, the remaining nontrivial conditions that must be verified are $\bra{i_L}S_X^2\ket{i_L}=\bra{j_L}S_X^2\ket{j_L}$ and $\bra{i_L}S_Y^2\ket{i_L}=\bra{j_L}S_Y^2\ket{j_L}$. 

In the case of $S_X^2$ projection, since $S_X$ can be expressed as $S_X=\frac{1}{2}(S_+ + S_-)$, the equation can be rewritten as  $\bra{i_L}(S_++S_-)(S_++S_-)\ket{i_L}=\bra{j_L}(S_++S_-)(S_++S_-)\ket{j_L}$. 

Due to the spacing between the $S_Z$-basis states, the only remaining nontrivial terms are $\bra{i_L}(S_+S_- + S_-S_+)\ket{i_L}=\bra{j_L}(S_+S_- + S_-S_+)\ket{j_L}$. 

Using the relations for the spin ladder operators, 
\begin{equation}
\begin{split}
S_+\ket{S,m}=\sqrt{S(S+1)-m(m+1)}\ket{S,m+1}\\
S_-\ket{S,m}=\sqrt{S(S+1)-m(m-1)}\ket{S,m-1},
\label{spin_ladder_operator}
\end{split}
\end{equation}

If we decompose $\ket{i_L}=\sum{c_m\ket{S,m}}$, the above term can be written as,

\begin{equation}
\begin{split}
\bra{i_L}(S_+S_- + S_-S_+)\ket{i_L}= 2S(S+1)-2\sum{\bra{S,m}m^2\lvert c_m \rvert^2 \ket{S,m}}=2S(S+1)-2\bra{i_L}S_Z^2\ket{i_L}.
\label{B6}
\end{split}
\end{equation}

Therefore, the conditions $\bra{i_L}S_X^2\ket{i_L}=\bra{j_L}S_X^2\ket{j_L}$ and $\bra{i_L}S_Y^2\ket{i_L}=\bra{j_L}S_Y^2\ket{j_L}$ also hold with the same coefficients as in Eq.\ref{qutrit_XYZ_correction_codeword}. The same reasoning can be iteratively extended arbitrary $2t+1$-distance codes, up to $\bra{i_L}S_X^{2t}\ket{i_L}=\bra{j_L}S_X^{2t}\ket{j_L}$ and $\bra{i_L}S_Y^{2t}\ket{i_L}=\bra{j_L}S_Y^{2t}\ket{j_L}$.


\section{Comparison of required Hilbert space dimension for encoding}
\label{SI_chapter_C}

\begin{figure}
\includegraphics[width=16cm]{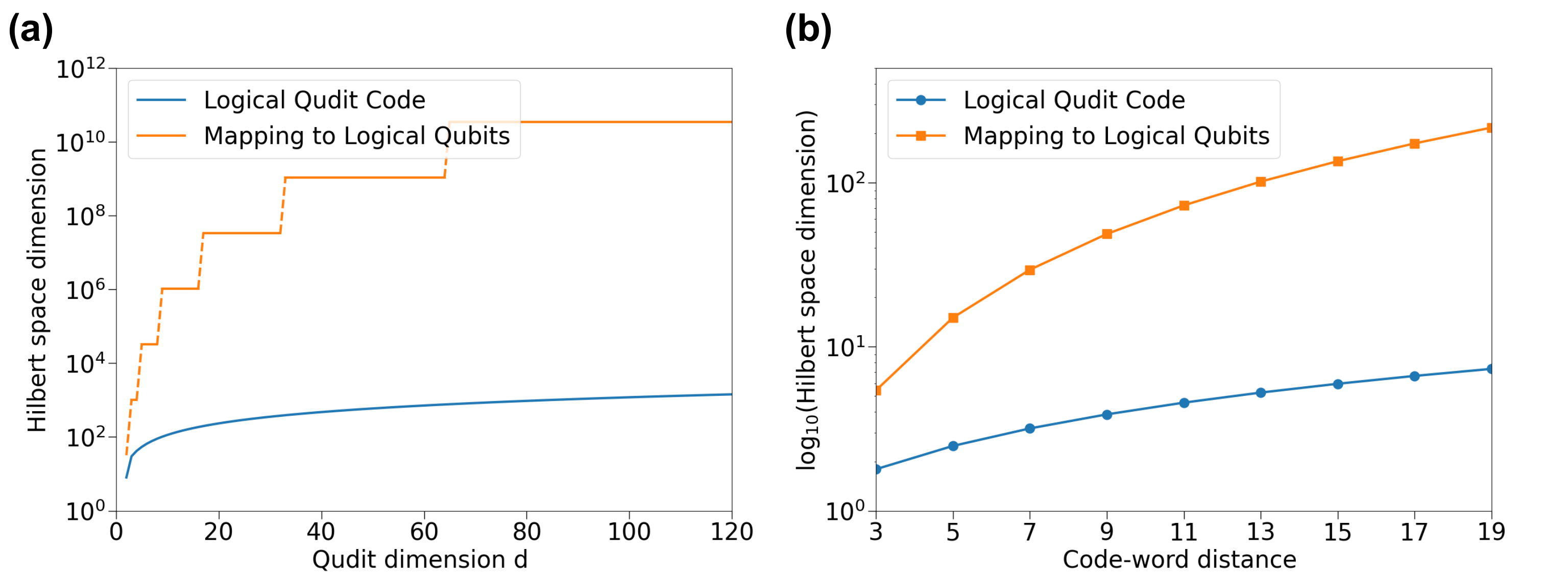}
\caption{Required Hilbert space for logical qudit encoding in this work (blue) and convensional mapping into multiple logical qubits (orange), according qudit dimension $d$ (a) and code-word distance (b). 
\label{SI_chapter_C_fig}}
\end{figure}

In Fig.~\ref{SI_chapter_C_fig}, we present the required Hilbert space for two strategies to realize a logical qudit: 1) logical qudit encoding scheme proposed in this work, and 2) conventional construction of a logical qudit from multiple logical qubits.  Fig.~\ref{SI_chapter_C_fig} (a) shows the scaling of Hilbert space dimension with qudit dimension $d$ for both approaches. As a working example, we consider the distance-3 logical qudit encoding described in Eq.~\ref{general_XYZ_correction_codeword}. For the conventional strategy, we assume that the logical qudit is constructed from multiple logical qubits, each encoded using a distance-3 surface code. Fig.~\ref{SI_chapter_C_fig} (b), shows the scaling of the Hilbert space dimension of logical qutrit as a function of code-word distance for the two strategies. 

As shown in both comparisons, encoding of logical qudits introduced in this work requires orders of magnitude smaller total Hilbert space dimension, implying that fault-tolerant encoding can be performed in a more resource-efficient manner than the conventional approach. The total Hilbert space dimension provides an upper bound on the physical error subspace that must be addressed during the error correction, and is relevant to experimental complexity may arise in practical implementations. While qubit-based QEC schemes nowadays face rapidly increasing decoding complexity with system size, and significant efforts have been devoted to developing compact and efficient decoding schemes\cite{demarti2024decoding}, such approaches generally include trade-offs between decoding scaling and error threshold. From this perspective, reducing the required Hilbert space offers a complementary and physically meaningful advantage.

\end{document}